\documentclass[journal,web]{ieeecolor}

\usepackage{tmi}
\usepackage{cite}
\usepackage{url}
\usepackage{amsmath,amssymb,amsfonts}
\usepackage{algorithmic}
\usepackage{graphicx}
\usepackage{textcomp}

\usepackage{amsfonts}
\usepackage{multicol}
\usepackage{float}
\usepackage{epstopdf}	
\usepackage{longtable}
\usepackage{comment}
\usepackage{xspace}
\usepackage{transparent}
\usepackage{tabularx,ragged2e,booktabs}
\usepackage{nomencl} 

\newtheorem{theorem}{Theorem}

\newtheorem{assumption}{Assumption}
\newcommand{\todo}[1]{\textcolor{blue}{[TODO] #1}}
\def\BibTeX{{\rm B\kern-.05em{\sc i\kern-.025em b}\kern-.08em
    T\kern-.1667em\lower.7ex\hbox{E}\kern-.125emX}}
\begin{document}
\title{Surgical Foundation Model Leveraging Compression and Entropy Maximization for Image-Guided Surgical Assistance}
\author{
Lianhao Yin$^{1,2}$, 
Ozanan Meireles$^{1,4}$, 
Guy Rosman$^{1,4}$,
Daniela Rus$^{2}$
\thanks{$^{1}$Surgical Artificial Intelligence Laboratory, Massachusetts General Hospital, MA, USA. {\tt\small lyin5@mgh.harvard.edu}}%
\thanks{$^{2}$Computer Science and Artificial Intelligence Laboratory, Massachusetts Institute of Technology, MA, USA. {\tt\small lianhao,rus@csail.mit.edu}}%
\thanks{$^{4}$Department of Surgery, Duke University, NC, USA. {\tt\small guy.rosman,ozanan.meireles@duke.edu}}%
\thanks{Acknowledgement: The authors receive partial research support from the Department of Surgery at Massachusetts General Hospital as well as CRICO}
\thanks{L.Y is the first author. Authors sequence in alphabetical order}
}
\maketitle

\begin{abstract}

Real-time video understanding is critical to guide procedures in minimally invasive surgery (MIS). However, supervised learning approaches require large, annotated datasets that are scarce due to annotation efforts that are prohibitive, e.g., in medical fields. Although self-supervision methods can address such limitations, current self-supervised methods often fail to capture structural and physical information in a form that generalizes across tasks. We propose Compress-to-Explore (C2E), a novel self-supervised framework that leverages Kolmogorov complexity to learn compact, informative representations from surgical videos. C2E uses entropy-maximizing decoders to compress images while preserving clinically relevant details, improving encoder performance without labeled data. Trained on large-scale unlabeled surgical datasets, C2E demonstrates strong generalization across a variety of surgical ML tasks, such as workflow classification, tool-tissue interaction classification, segmentation, and diagnosis tasks, providing improved performance as a surgical visual foundation model. As we further show in the paper, the model's internal compact representation better disentangles features from different structural parts of images. 
The resulting performance improvements highlight the yet untapped potential of self-supervised learning to enhance surgical AI and improve outcomes in MIS.
\end{abstract}


\section{Introduction}
Artificial intelligence-enabled video analysis and minimally invasive surgery (MIS) are revolutionizing surgical practices, enhancing patient outcomes in multiple tasks throughout the preoperative, intraoperative, and postoperative surgical lifecycle. AI algorithms significantly improve preoperative planning by accurately segmenting medical images, reducing surgical errors, and optimizing resource allocation \cite{oh2023automated,chen2022ai}. In the surgical intraoperative setting, computer vision approaches can help evaluate the difficulty of surgery, guide safe dissection, improve staff awareness and readiness in real time by detecting tools, tissues, workflow stages, and performing tissue segmentation~\cite{Haouchine13,mascagni2022computer,Zhang-uubpu}. 
The use of AI-assisted techniques in surgical practices has advantages such as reduced postoperative complications, shorter hospital stays, and improved patient prognoses \cite{Morris2024-jb, Ashrafiana2024-vx}.

The videos captured by the camera during minimally invasive surgery serve as the main input for minimally invasive surgery, providing critical information for better decision-making and better care \cite{Avram2023-zy}. Traditionally, enabling machines to understand the content of these image sequences and provide assistance to surgeons during or after surgical procedures requires a high quality and quantity of image labeling and supervised learning methods to train. In supervised learning, the precision of the predictions correlates with the number of labeled data based on Hoeffding’s inequality and the union bound\cite{shalev2014understanding}.
In contrast, humans learn new visual concepts with far fewer labeled examples, in a manner that is akin to self-supervised learning approaches, which rely significantly on the input data examples and only leverage a smaller number of labels in the final training stages for specific tasks. Such approaches are important in the medical domain, especially when we aim to extend surgical AI to new and uncharted surgical procedures, where labeled datasets are often unavailable, highly imbalanced, or restricted due to privacy concerns and infrastructure limitations \cite{Meireles2021-nx,Eckhoff2023-pg,Avram2023-zy}. 
In addition to the issues related to limited high quality and quantity of training data in surgery, the taxonomies used in surgical and medical datasets are often narrower compared to the breadth of labels captured in large computer vision datasets \cite{deng2010does}. 

Surgical images and videos, as well as other medical images, are rich in information, containing abundant local features critical to downstream tasks \cite{Gui2024-gh, Wang2023-nc,Sabottke2020-sh}, which makes it even more important for the balance of global and local information with compact representations. For instance, distinguishing between organs such as the liver and the heart often requires not only local information, such as textures and colors, but also global information, such as shapes, similar to how one distinguishes a bird from a dog in ImageNet. Textures and local features are particularly crucial in medical images, especially for diagnostic tasks. For example, detecting fine microcalcifications and details is critical for identifying breast cancers \cite{Rangarajan2022-dt}. Similarly, the polyps in the colon vary, and micropolyps are identifiable only with changes in local texture, color, or shape. In summary, surgical image analysis is based on fine-grained phenomena~\cite{Sabottke2020-sh}, we are still missing self-supervised approaches that can handle such perceptually detailed data while leveraging the limited number of concepts that are used to reason about each section of the video.

\begin{figure*}[ht]
    \centering
    \includegraphics[width=0.97\linewidth]{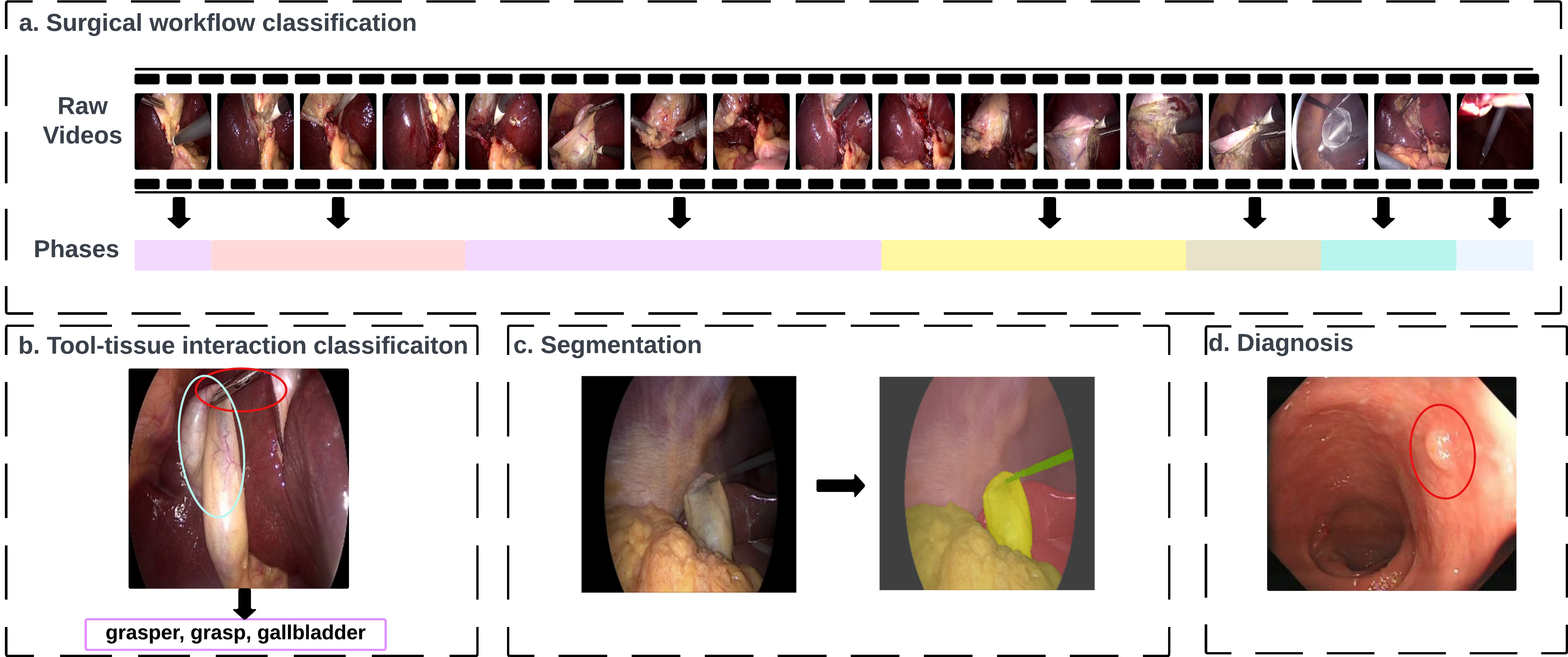}
      \caption{The generalization of the pretrained encoder was evaluated across various downstream tasks, such as (a) workflow classification, (b) action classification, (c) segmentation, and (d) diagnosis}
    \label{fig: finetuningtask}
\end{figure*}

We propose a self-supervised learning method that uses Kolmogorov complexity and entropy maximization methods to optimally compress information from local and global views with compact representations, encompassing both semantic and perceptual information and to overcome the bias introduced in the self-supervised training process, such as the Masked Auto Encoder (MAE)~\cite{He2022-la}, to achieve balanced performance across different sets of downstream tasks. The majority of current computer vision approaches in surgery use visual feature backbones that were pre-trained from general images, which is bias from the surgical images. Some works have used the standard MAE method to train with videos of general surgery and demonstrated results in limited tasks such as phase classification~\cite{jaspers2024exploring} and segmentation and limited surgeries such as laparoscopic cholecystectomy~\cite{batic2024endovit}. We pre-train the foundation model using the largest surgical image datasets combining both public and private available datasets from Massachusetts General Hospital (MGH) using the proposed self-supervised learning method. The results demonstrate promising outcomes and generalization capability across various downstream tasks using a pre-trained neural network. Downstream tasks can include surgical workflow classification, surgical action classification, segmentation, and diagnosis (Fig.~\ref{fig: finetuningtask}), which will be detailed in Section~\ref{sec_results}. 
We show that the resulting latent compressed visual model, Compress-to-Explore (C2E), provides an effective visual foundation model for surgical video analysis, while catering to both local (perceptual) and global (semantic) aspects of the images~\cite{Rombach2022-ic}. The contributions of the paper are as follows.
\begin{itemize}
    \item  We devise the latent compression transformer architecture and define how to train it to obtain latent compressed representation within the auto-encoder architecture.
    \item We demonstrated the generalization capability of the proposed self-supervised learning foundation model in various downstream task, compact representation, and few-short learning capabilities.
    \item We demonstrate how the approach allows us to expand on a novel dataset with 0.78M images based on 2122 surgeries (845 surgeries collected over 8 years from Massachusset General Hospital and other 1277 surgeries from open datasets), demonstrating how the approach enables the scaling up of surgical foundation model.
\end{itemize}

\subsection{Related work}

\subsubsection{Self-supervised foundation models}
Self-supervised foundation models~\cite{bommasani2021opportunities} have been changing multiple domains of AI and many aspects of society. Foundation models are pretrained using self-supervised method and finetuned in various of downstream tasks. 
Two main branches of self-supervised learning have been widely researched in computer vision: invariance-based methods and generative methods. 

\textbf{Invariance-based} One of the examples of invariance-based methods is contrastive-based learning such as Simple Siamese (SimSiam) \cite{Chen2020-jq, Hirsch2023-vl}. It backpropagates the error through the image encoder by maximizing the similarity of two images augmented from the same image. Some contrastive learning has been applied for pretraining in surgical video analysis \cite{Yu2022-ac,yengera2018less,neimark2021train,ramesh2023dissecting}. Another example is the Joint Embedding Predictive Architecture (JEPA) \cite{Assran2023-wk}, which predicts the embedding of one part of the image using the embedding of the other parts of the same image. Invariance-based self-supervised learning requires balancing the learning of local and global features to optimize performance across downstream tasks, such as classification and segmentation, where, the classification task requires more global features than local features compared to the segmentation task \cite{Bardes2022-es}. To address this challenge, several approaches have been proposed, such as using teacher-student networks to capture local and global features \cite{Wang2023-nc}.

\textbf{Generative model} The widely used generative methods in surgical video analysis are the variational auto encoder (VAE), the masked auto encoder (MAE) \cite{He2022-la}, the diffusion models \cite{Ho2020-ef}, etc. 
The MAE pre-trained encoders were used in various downstream tasks by predicting masked images patches using unmasked images patches, which has shown promise in generalization \cite{batic2024endovit} and will be adopted in this paper.  
Diffusion-based models are often incorporated in more specific tasks such as segmentation. Such diffusion models have also been used to address the imbalance of training data by generating synthetic images for underrepresented scenarios \cite{Kazerouni2022-gk}. 



\subsubsection{Information-based approaches}
Information-based approaches have been used in computer vision \cite{jain1976fast} and deep neural network design \cite{ma2007segmentation,arora2014provable,yu2023white}. Recent works have broadly looked at effective but compact representations~\cite{tishby2015deep,Elad19,Goldfeld2020,Zhaoyang23}. For example, total coding rate maximization has been used as covariance regularization techniques to avoid collapse of representation, and coding rate reduction inspired design has improved the representation of self-supervised learning \cite{li2022neural,liu2022self}, segment images \cite{Ma2007-eo} and design neural network \cite{yu2023white}. The compression of neural network input using the discrete cosine transform (DCT), which is an image compilation method, was shown to be 5 times faster in video language action training and improved evaluation score \cite{pertsch2025fast}. This paper proposed a new of information compression leverage Kolmogorov complexity minimization to be explained in Section.~\ref{se: compression}.

\subsubsection{Energy-based approaches}
Compositionality enables better generalization \cite{lake2023human} capacity of neural networks. Several attempts are made to add an energy-based model to the generative model framework work \cite{du2021unsupervised} to enable compositional understanding of the scene \cite{gkanatsios2023energy} and better generalization \cite{du2020compositional}. We adapted the closed-formed model of entropy maximization which represents a general energy-based model formation, and used Langevin sampling of such closed-formed entropy maximization in the decoder design to provide a better compositional and generalizable capability, which will be explained in Section.~\ref{se: exploration process}.
\subsubsection{Surgical Computer Vision - Overview and Self-Supervised Learning}
Finally we demonstrate our results in the domain of surgical computer vision. Surgical computer vision has been extending what AI can understand and act on in the surgical context \cite{Hashimoto2018-vd}, with a variety of efforts aimed at segmentation~\cite{Bodenstedt2016,Garcia-Peraza-Herrera2017-kq,Allan18Robotic,hong2020cholecseg8k}, detection~\cite{Sarikaya2017-ch,Jin2018-jj,Kondo2021-jv,bouget2017vision}, tracking~\cite{speidel2006tracking,zhao2017tracking}, workflow estimation~\cite{padoy2012statistical}, and prediction~\cite{jalal2019predicting, ban2022supr,yin2024hypergraph}.
As we pointed out in the introduction, while much of the part work in the field depended on supervised learning or the existence of large labeled datasets, visual foundation models~\cite{Gui2024-gh,Wang2023-nc,liu2024surgical,batic2024endovit} are changing the fields by enabling both superior results as well as forays into new procedures and taxonomies.

\begin{figure*}[!ht]
    \centering
    \includegraphics[width=0.8\linewidth]{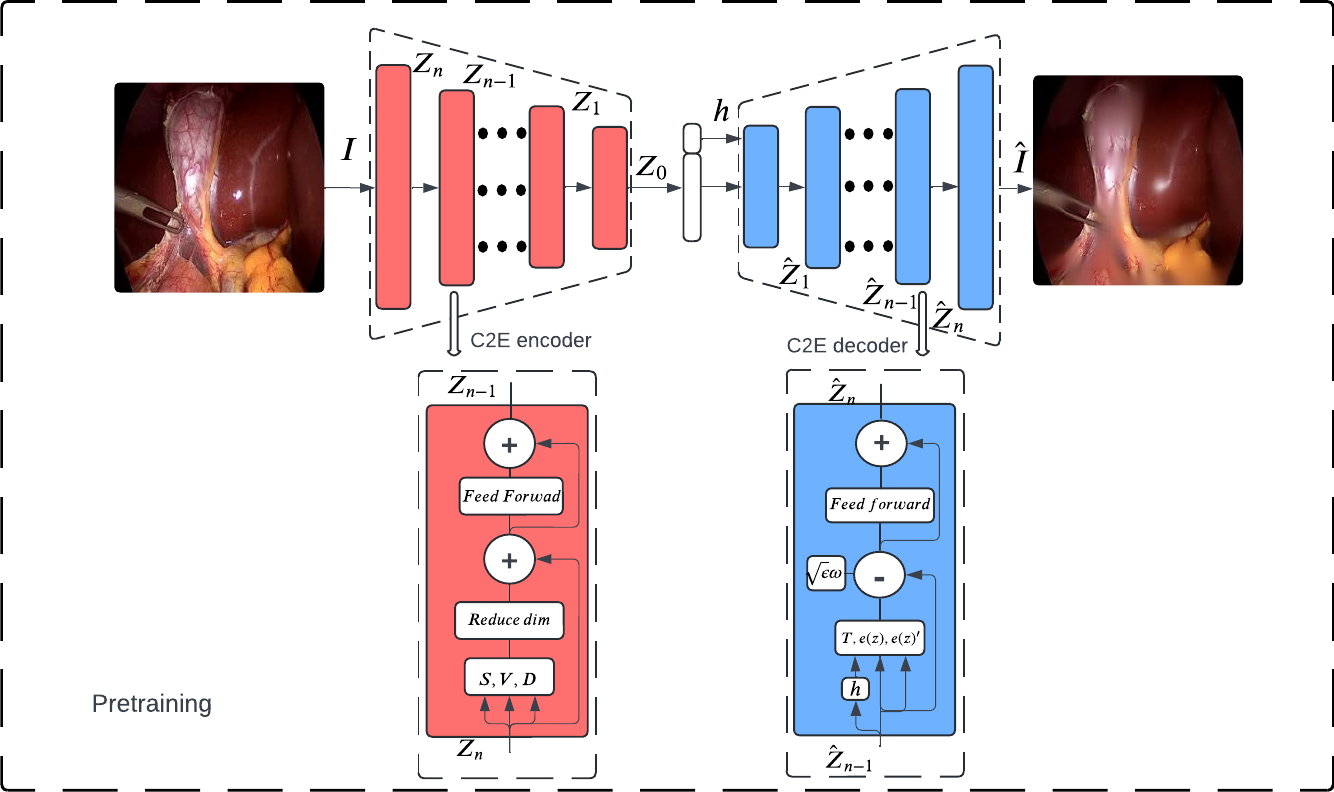} 
    \caption{C2E General structure. In pretraining, the encoder compresses the raw images patches $I$ into hidden state $Z_0$ using $n$ layers of proposed encoder structure.The output of hidden state of each layer is named as $Z_{i}$, where $i \in[0,n]$. The decoder reconstruct the images patches and the the output of each layer are as estimate of hidden state $\hat{Z}_{i}$, where $i\in[0,n]$. The error of the reconstructed images and the raw images are backpropogated to the neural network during the training process. In pretrained encoder were used to extract features from the videos and used for a few downstream tasks (Fig.~\ref{fig: finetuningtask}) such as surgical work flow prediction, surgeons' action prediction, segmentation, diagnosis.}
    \label{fig:c2e structure details}
\end{figure*}

\section{Methods} \label{section: method}

We describe our self-supervised model and training approach in this section. 
The neural network adopted the MAE structure to use an unmasked part of the image to predict the masked part of the images, using an auto-encoder structure (see Fig.~\ref{fig:c2e structure details}). The masking rule is inherited from the MAE structure. We changed the structure of the encoder and decoder accordingly with a compressor encoder and explorer decoder using proposed compression and entropy maximization approaches that will be explained in Section \ref{se: compression} and Section \ref{se: exploration process}. 

\subsection{Notation}
The images represented by $I \in \mathbb{R}^{W_{img}\times H_{img} \times 3}$ are the input training data, where $W_{img}$ is the width and $H_{img}$ is the height of the images. The compression step compresses the input $I$ through $n$ steps and the hidden layer of each step is denoted $Z_n$. The hidden state of the final compression step is $Z_0$, $h$ is the hidden state of the temperature measurement within $Z_0$ that can represent the temperature level of the inputs. Kolmogorov complexity and the complexity of the hidden state are denoted by $K(Z)$, and the entropy of that is defined by $H(Z)$. The $\hat{Z}$ represents an estimation of the hidden state in the decoder. The energy function uses $E$ to represent the energy of the hidden state.  

\subsection{Compression encoder} \label{se: compression}
We want to compress the input $I$ to the hidden space $Z_0 \in R^{N,C}$, which is a disentangled and sparse normal distribution, where $N$ is the number of patches and $C$ is the number of elements. Based on \cite{Cover2006-se} (Theorem 8.4.1), entropy of multi-variable normal distribution is 
\begin{equation} \label{eq: entropy of multivariable Gaussion}
h(Z_1, Z_2, ..., Z_n ) = h(N(\mu,\Sigma) = \frac{1}{2}\ln{(2\pi e)^M|\Sigma|}
\end{equation}
And it can be expanded as Eq.~\ref{eq: Kolmogorov complexity}.
\begin{equation} \label{eq: Kolmogorov complexity}
\begin{aligned}
    H(Z) = \frac{1}{2} \ln{|\Sigma|} + \frac{M}{2} (1 + \ln{2\pi})
\end{aligned} 
\end{equation}
where $M$ is the dimensionality of $x$. The covariance matrix can be approximated by $\Sigma = \frac{1}{m}ZZ^T$, with $Z$ having a mean of zero due to a normalization layer. 

We want to compress $I$ to the hidden state $Z_0$ with a high compression ratio. We formulate this as maximizing the Kolmogorov complexity difference ($K(Z)-K(I)$) for each compression step.  
Theory \ref{theorem: min H min Kolmogorov complexity} has shown that $ \mathbb{E} \frac{1}{n}K(Z^n|n) \rightarrow H(Z)$, which means that the Kolmogorov complexity approaches the entropy defined in \ref{eq: Kolmogorov complexity}. Therefore, for the compression process, maximizing the Kolmogorov complexity difference is equal to maximizing the entropy difference of the input and hidden states as in Eq.~\ref{eq: maximize compression = maximize entropy difference}, with the dimensionality constraints in the equation. Theorem~\ref{theorem: convex of maximize compression} indicates that Eq.~\ref{eq: maximize compression = maximize entropy difference} is concave and the solution is located at the vertices of feasible sets of hidden states.
\begin{equation} \label{eq: maximize compression = maximize entropy difference}
    \begin{aligned}
        \max_{\theta} \quad & - H(Z) + H(I) \\
           \propto & - \frac{1}{2}(\ln{|\Sigma_{Z}|} - \ln{|\Sigma_{I}|}) + \frac{1}{2}(\Delta M) \\
           \propto & - \frac{1}{2}(\ln{|\Sigma_{Z}|}) + \frac{1}{2}(\Delta M) 
    \end{aligned}
\end{equation}
 If we assume that the dimension changes are constant, $\frac{1}{2}\Delta M$ will have no influence on the optimization. We also prove that the optimization (Eq.~\ref{eq: maximize compression = maximize entropy difference}) holds with the hidden state with reduced dimension in Theorem~\ref{theorem: dimension reduction process}. In neural network design, before each encoder block project, we remove a constant dimension to enforce the dimension reduction (Fig.~\ref{fig:c2e structure details}). 

We then use iterative shrinkage-thresholding algorithms to solve the optimization (Eq.~\ref{eq: maximize compression = maximize entropy difference}) in an iterative way, where the step index $i\in[0,n-1]$. Taking $Z = SVD$, the optimization in Eq.~\ref{eq: maximize compression = maximize entropy difference} can be split into multiple compression steps to compress $Z_n$ to $Z_0$ in Eq.~\ref{eq: z = z_k -(-H + H)}, which minimize the Kolmogorov complexity proved by Theorem.~\ref{theorem:  z = z_k -(-H + H) minimize Kolmogorov complexity}. 
\begin{equation} \label{eq: z = z_k -(-H + H)}
    \begin{aligned}
        Z_{i+\frac{1}{2}} = & Z_{i+1} + \beta \nabla_{Z}(-H(Z_{i+1}))    \\ 
          = & Z_{i+1} - \beta Z_{i+1} (Z_{i+1}^T Z_{i+1})^{-1}                          \\
          = & Z_{i+1} - \beta (S_{i+1}V_{i+1}D_{i+1})\\
          &(D_{i+1}^T V_{i+1} S_{i+1}^T S_{i+1} V_{i+1} D_{i+1})^{-1}                  \\
          = & Z_{i+1} - \beta (S_{i+1} V_{i+1}^{-1} D_{i+1})                            \\
    \end{aligned}
\end{equation}
In the end, we applied the linear feedforward layer to project $Z_{i+\frac{1}{2}}$ to subspaces $Z_i$ at each iteration in Eq.~\ref{eq: subspace iteration} with a bypass term. The bypass term is to enable information and gradient propagation through multiple layers of deep neural network as residual function used by ResNet~\cite{he2016deep}.  
\begin{equation} \label{eq: subspace iteration}
    Z_i = Z_{i+\frac{1}{2}} + PZ_{i+\frac{1}{2}}
\end{equation}
Based on Theorem~\ref{theorem: maping Z to subpsace maximize the compression}, mapping $Z_{i+\frac{1}{2}}$ to subspaces where $P^TP = \Pi$ maximize the compression, where, $\Pi \in R^{C,C}$ is a diagonal matrix with $\pi_i \in [0,1]$. The decoder used inverse of such mapping.  
We note that in our implementation, we use $S,V,D,P$, conditioned on the input to each layer, as linear operators, without constraining them into their specific matrix manifold \cite{rosman2014fast}. 

\subsection{Exploration decoder} \label{se: exploration process}
The motivation of an exploration decoder is that an inverse compression process is to maximize entropy to reconstruct the images. The solutions of entropy maximization in a closed-formed solution are in energy functions formulation, whereas the energy-based neural network in the decoder design has the potential to have a better capability of generalization. The other motivation is that the entropy maximization process itself can also improve the representation and compression process. 

From the maximization of entropy with the expectation of an energy function $E(z)$ as a neural network, where $z$ is the sample of elements of $Z$, the probability of state at the energy of $E(z)$ is $p_k$ which is the solution of maximization of Eq.~\ref{eq: Boltzmann energy maximization}~\cite{Jaynes1957-hu} with Lagrange multipliers.  
\begin{equation} \label{eq: Boltzmann energy maximization}
    \begin{aligned}
        J  =& -K \sum_{k=1}^{n} p_k \log(p_k)  + \lambda ( 1- \sum_{i=1}^{n} p_k )  \\ 
            & + \mu ( \mathbb{E}[E(z)] - \sum_{k=1}^{n}p_{k}E(z_k) )  \\
    \end{aligned}
\end{equation}
The probability of the state of an individual particle is a function of the expectation of function $E(z)$. Taking $E(z)$ as energy, we can get Eq.~\ref{eq: bio: p boltzmann} as a closed-form solution of Eq.~\ref{eq: Boltzmann energy maximization}. 

\begin{equation} 
    \begin{aligned}
        p(z) \propto e^{-\frac{1}{\mathbb{E}[E(z)]}E(z)}
    \end{aligned}
    \label{eq: bio: p boltzmann}
\end{equation}

The samples of $z$ are conditioned on the expectation of energy $\mathbb{E}[E(z)]$ that is proportional to the temperature $kT(h)$, where the temperature can be represented as a function of a hidden state $h$.
\begin{equation}
    \begin{aligned}
        \mathbb{E}[E(z)] = kT(h)
    \end{aligned}
\end{equation}

With Langevin dynamics, the hidden state can be sampled by following steps:
\begin{equation} \label{eq: langevin sampling}
\hat{Z}_{i} = \hat{Z}_{i-1} + \epsilon \nabla \log{p(\hat{Z}_i)}  + \sqrt{2 \epsilon} \omega_i \,\, \omega_i \sim N(0, I) \\ 
\end{equation}
where $N$ is a multivariate Gaussian, and $i \in [0, n]$. 

Plug in Eq.~\ref{eq: bio: p boltzmann} to Eq.~\ref{eq: langevin sampling}, and combine an inverse process of Eq.~\ref{eq: subspace iteration}, we can estimate the hidden state in iterative way Eq.~\ref{eq: Z_k-1/2 = z_k-1},~\ref{eq: inverse subspace iteration}, which is also illustrated in Fig.~\ref{fig:c2e structure details}.  
\begin{equation} \label{eq: Z_k-1/2 = z_k-1}
\begin{aligned}
\hat{Z}_{k-\frac{1}{2}}  & = \hat{Z}_{k-1} - \epsilon\frac{1}{kT(h)}E(\hat{Z}_{k-1})E(\hat{Z}_{k-1})' + \sqrt{2 \epsilon} \omega_i 
\end{aligned}
\end{equation}
\begin{equation} \label{eq: inverse subspace iteration}
    \hat{Z}_{i} = \hat{Z}_{i-\frac{1}{2}} + P^{-1}\hat{Z}_{i-\frac{1}{2}}
\end{equation}
Since the data are limited and the samples in each subspace of $Z_0$ are normally distributed, the temperature can be estimated by part of the subspace within which the hidden states are uniformly distributed. 
The temperature of the exploration samples is conditioned by partial elements of $Z_0$, which is proved by Theorem \ref{theorem: temperature sub}. In the implementation, we proposed a conditional ratio $\beta \in [0,1]$ to enforce partial elements of hidden stats to represent the total hidden states. 
\begin{equation} \label{eq: epsilon = f(Z_0)}
h = E_c(Z_0,\beta)
\end{equation}

\subsection{Pre-training} \label{section: pretraining datasets}
\label{sec_results}
For pretraining, we used videos from various open source surgical videos including laparoscopic cholecystectomy, proctocolectomy, rectal resection, sigmoid resection, gynecologic laparoscopy, radical prostatectomy, sleeve gastrectomy, and the largest surgical video datasets in general surgery from Massachusetts General Hospital. We take 1 fps frame extracted from the raw videos and filtered the images by deduplicating them \cite{idealods2019imagededup}, and then using CLIP-based image quality assessment \cite{wang2023exploring} to keep only distinct and high-quality images, to have a dataset with largest image diversity in surgery. 

We also remove the images that were used for downstream tasks to avoid data leakage. The resulting datasets for pretraining include 2122 surgeries and 0.78M images (Tab.~\ref{tb: surgical datasets sources}). 

As stated in Theorem \ref{theorem: min H min Kolmogorov complexity}, the objective of the compression process is convex and differentiable. The loss function minimizes the mean average error of the predicted images ($\hat{I}$) and the original images ($I$), making the overall training run as a standard MAE~\cite{He2022-la} pipeline.

\begin{table}[ht]
    \centering
    \caption{Pretraining datasets}
    \begin{tabular}{c|c|c|c}
        \hline
        Surgical                                & Surgical                  & Surgery                 & Unique   \\ 
        categories                              & datasets                  & numbers                 & images \\\hline
        Lap Chole                               & Cholec80                  & 80                      & 30000   \cite{Twinanda2017-sj}\\ 
                                                & hSDB instrument           & 24                      & 10000   \cite{yoon2021hsdb}\\
        Proctocolectomy                         & HeiCo                     & 10                      & 14448   \cite{maier2021heidelberg} \\
        Rectal Resection                        & HeiCo                     & 10                      & 19528\\
                                                & DSAD                      & 32                      & 4239     \cite{carstens2023dresden}\\
        Sigmoid Resection                       & HeiCo                     & 10                      & 16037\\
        Gynecologic laparoscopy                 & LapGyn4                   & 500                     & 10000  \cite{leibetseder2018lapgyn4}\\
                                                & SurgicalActions160        & 59                      & 491 \cite{schoeffmann2018video}\\
                                                & GLENDA                    & $>$ 400                 & 578 \cite{leibetseder2019glenda}\\ 
        Radical prostatectomy                   & ESAD                      & 4                       & 5768 \cite{bawa2021saras}\\
        Sleeve gastrectomy                      & MGH                       & 86                      & 135766 \\ 
        General surgery                         & MGH18k                    & 845                     & 531565 \\ 
        \hline
        Total                                   &                           & 2122                    & 788420 \\
        \hline
    \end{tabular}
    \label{tb: surgical datasets sources}
\end{table}

\section{Results}\label{sec_results}
We proceed to explore the performance of our model.
We demonstrate that the proposed method has a strong generalization capacity as a pre-trained encoder in various downstream tasks and in various surgery types, thus functioning as a visual foundation model for surgery.
We then proceed to analyze the intermediate representations afforded by the model, which demonstrate its superior ability to disentangle known medical concepts such as different tools, surgical phases, and type of procedure.

For quantitative results on downstream tasks, we examine the model's few-shot behavior across a variety of procedures and ML tasks in Section~\ref{section: resutls, analysis, tsne, few-short}. We focus on standard fine-tuning ML tasks in Sections~\ref{section: results, workflow prediction}-\ref{section: results, polyps}.  
For each of the downstream tasks, we connect the pre-trained encoder to downstream-task decoder (Fig.~\ref{fig:c2e structure details}) for fine-tuning, demonstrating how the model functions as a surgical foundation model and enables task generalization. To provide a fair comparison, we selected related work using base size of encoder which is similar to our encoder size (85.22M parameters for encoder) to contrast in the results. The pre-trained encoder also demonstrated better few-shot learning capability in out-of-distribution datasets in Section~\ref{section: resutls, analysis, tsne, few-short}. 

For qualitative analysis, to provide more insights of neural network representation, we also analyzed the representation and saliency maps of various tasks in Section~\ref{section: tsne}. 

\subsection{Surgical workflow prediction} \label{section: results, workflow prediction}
Surgery typically progresses through distinct phases during treatment. Automatically understanding and analyzing the surgical workflow can yield valuable insights, such as detecting abnormalities such as unusually prolonged or shortened phase durations or identifying skipped phases.  These will provide warning and prediction of the next step for surgeons and therefore improve patient care. 
Specifically, workflow prediction can be formulated as a classification problem in the downstream task. 
We adopted Cholec80 datasets for the evaluation of the workflow prediction of cholecystecomty surgery. The Cholec80 datasets~\cite{Twinanda2017-sj} contains 80 cholecystectomy surgery videos, annotated with 7 distinct phase labels. We used the official splitting (40 videos for training and 40 videos for testing). 
The results show that the proposed method can improve the classification accuracy by replacing the TeCNO backbone in the downstream task (Tab.~\ref{tb: phase}). 
\begin{table}[ht]
    \centering
    \caption{SOTA in phase recognition with Cholec80}
    \begin{tabular}{ccc}
    \hline
    Method                                          & Accuracy   & Precession               \\ \hline  
    EndoNet~\cite{Twinanda2017-sj}                   & 81.7 ± 4.2 & 73.7 ± 16.1              \\ \hline 
    EndoNet+LSTM~\cite{Twinanda2017-go}              & 88.6 ± 9.6 & 84.4 ± 7.9               \\ \hline 
    MTRCNet-CL~\cite{Jin2020-wa}                     & 89.2 ± 7.6 & 86.9 ± 4.3               \\ \hline 
    PhaseNet~\cite{Twinanda2017-sj,Twinanda2016-na}  & 78.8 ± 4.7 & 71.3 ± 15.6              \\ \hline
    SV-RCNet~\cite{Jin2018-bw}                       & 85.3 ± 7.3 & 80.7 ± 7.0               \\ \hline
    TeCNO~\cite{Czempiel2020-hz}                     & 88.6 ± 7.8 & 86.5 ± 7.0               \\ \hline
    Trans-SVNet~\cite{Gao2021-sp}                    & 90.3 ± 7.1 & 90.7 ± 5.0               \\ \hline             
    LoViT~\cite{Liu2024-tz}                          & 91.5 ± 6.1 & 83.1± 9.3                 \\ \hline   
    SurgFormer ~\cite{yang2024surgformer}            & 92.4 ± 6.4 & 87.9 ± 6.9        \\ \hline
    \textbf{C2E(ours)}                              & \textbf{92.5±6.9}         & 83.2±13.9        \\ \hline 
    \end{tabular}
    \label{tb: phase} 
\end{table}

\subsection{Surgeon's action} \label{section: results, behavior and action}
Monitoring and forecasting surgeons' actions in real time provides detailed information on what tissues and what tools the surgeon is working on. This monitoring and forecasting will prevent problems such as surgeon cutting or resecting in wrong-target tissues or in the wrong phase of a surgery. In the end, this improves the safety and patient care. We used the evaluation metrics proposed for the CholecT45~\cite{Nwoye2023-ci} data set for the downstream task. We adopted Rendezvous method using focal loss and long videos as input without changing the decoder structure and only changing the backbones. The proposed method has better average precision in detecting taken, tissues, tool-tissue interaction, tool-action pair, and action-tool-tissue pairs (tab.~\ref{tb: Action triplet}). 

\begin{table}[ht]
    \centering
    \caption{Comparison of the proposed method with related work in action triplet}
    \begin{tabular}{ccccc}
    \hline
    Method                                  & $mAP_{it}$                    & $mAP_{iv}$               &  $mAP_{ivt}$          \\ \hline 
    Tripnet~\cite{Nwoye2020-qm}              & 27.1$\pm$2.8                  & 31.8$\pm$4.1             & 24.4$\pm$4.7          \\ \hline 
    Attention Triplet~\cite{Nwoye2022-ie}    & 29.4$\pm$1.2                  & 33.0$\pm$2.9             &  27.2$\pm$2.7         \\ \hline 
    Rendezvous~\cite{Nwoye2022-ie}          & 30.8$\pm$2.1                  & 34.0$\pm$3.3             & 29.4$\pm$2.8          \\ \hline 
    ConceptNet-ViT~\cite{Ban2023-rj}       & 33.4$\pm$3.2                  & 33.7$\pm$3.2             &  30.6$\pm$1.9         \\ \hline 
    RiT~\cite{Sharma2023-pl}                & 36.9$\pm$1.0                  & 38.3$\pm$3.5             & 29.7$\pm$2.6          \\ \hline 

    MT4MTL(No Ensemble)~\cite{Gui2024-gh}& 43.1$\pm$2.0                  & 44.9$\pm$2.4             & 37.1$\pm$0.5          \\ \hline 
    LAM~\cite{li2024parameter}                                     & -                             & -                        & 35.9                  \\ \hline   
    \textbf{ours}                           & 42.3$\pm$0.9                  & 48.6$\pm$2.1             & \textbf{41.8$\pm$0.9}  \\ \hline

    \end{tabular}
    \label{tb: Action triplet} 
\end{table}

\subsection{Segmentations in surgery images} \label{section: results, segmentation}
\begin{figure}[h!]
    \centering
\begin{minipage}{0.24\linewidth}
\centering
    \includegraphics[width=\linewidth,clip=true,trim = 0cm 15cm 42.75cm 0]{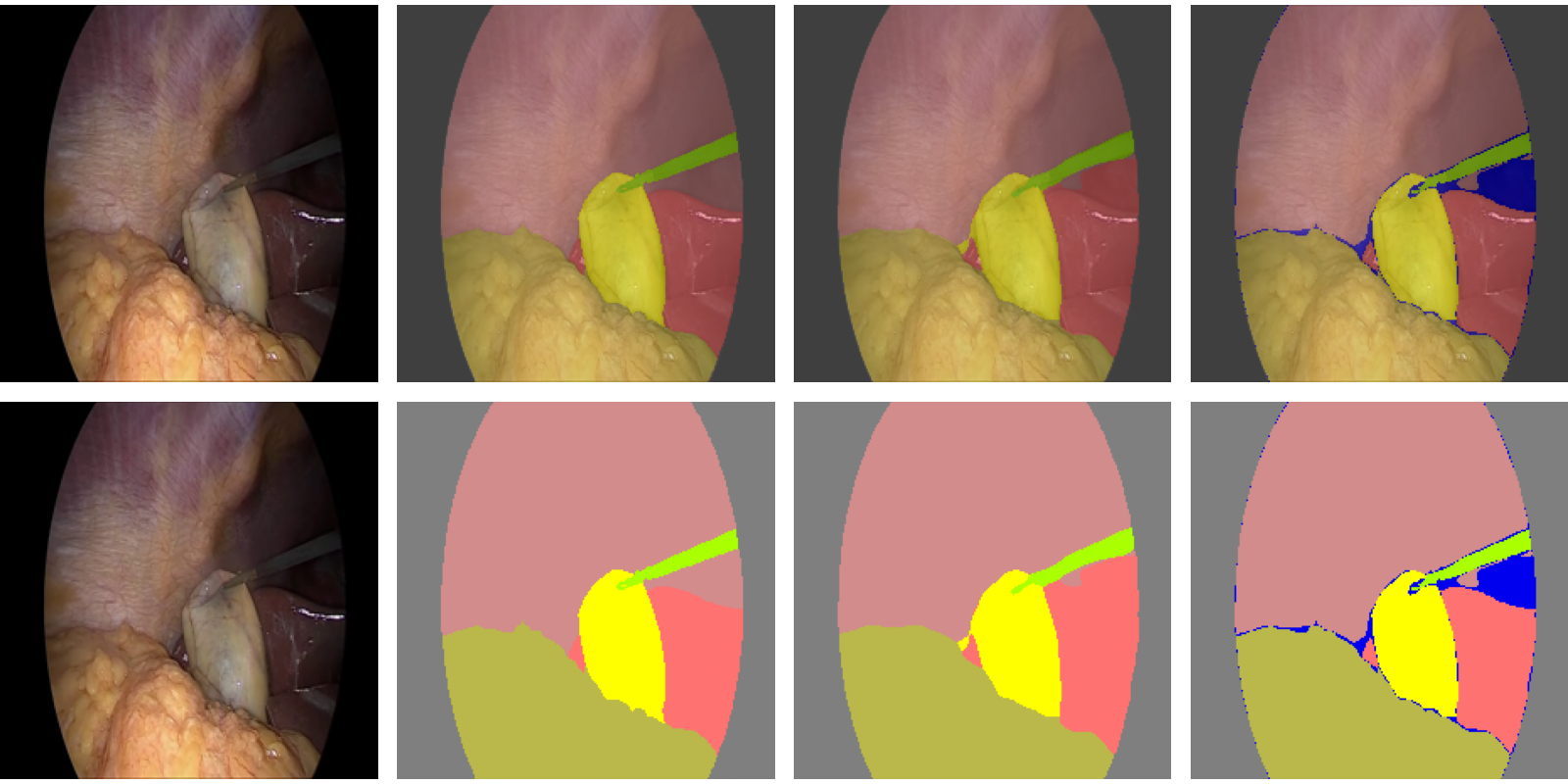}
a
\end{minipage}
\begin{minipage}{0.24\linewidth}
\centering
    \includegraphics[width=\linewidth,clip=true,trim = 14.25cm 15cm 28.5cm 0]{imgs/visualization/segmentation/mae_media_images_epoch=0799_val_Example_17_223_1dd1568cdb7ea54efd6a_1_.png}
b
\end{minipage}    
\begin{minipage}{0.24\linewidth}
\centering
    \includegraphics[width=\linewidth,clip=true,trim = 42.75cm 15cm 0cm 0]{imgs/visualization/segmentation/mae_media_images_epoch=0799_val_Example_17_223_1dd1568cdb7ea54efd6a_1_.png}
c
\end{minipage}  
\begin{minipage}{0.24\linewidth}
\centering
    \includegraphics[width=\linewidth,clip=true,trim = 42.75cm 15cm 0cm 0]{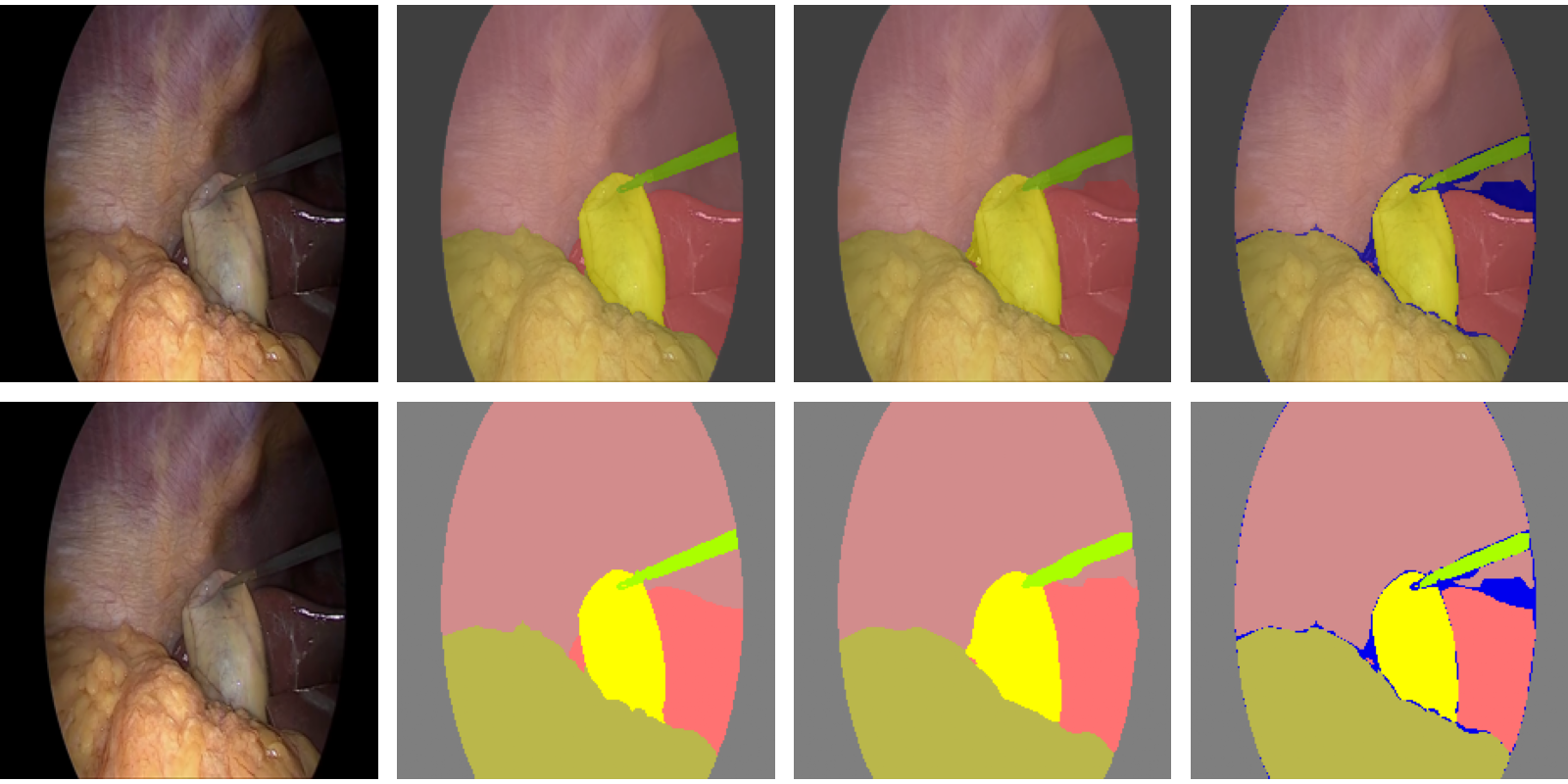}
d
\end{minipage}    
    \caption{Comparison of segmentation using VIT and C2E. a: The original input image. b: The ground truth of the segmentation. Subfigure c and d indicate the difference of ground truth of and prediction Segmentation using VIT(c) and C2E(d). The blue parts are the difference. Our proposed method has smaller difference (small blue areas) and has better semantic segmentation on the area between tool, liver and background }
    \label{fig: segmentation comparison of VIT C2E}
\end{figure}

Identifying distinguished anatomical structures is critical during surgery. A high accuracy of such task will improve the performance of the correct dissection, resection, and cutting. To have a fair comparison, we compared the proposed method with the method using the same fine-tuning structure. The proposed method has shown improved performance (see Tab.~\ref{tb: sota segmentation}) using the CholecSeg8k dataset~\cite{hong2020cholecseg8k}, adopting the downstream decoder from the Dense Prediction Transformer (DPT) decoder. A demonstration (Fig.~\ref{fig: segmentation comparison of VIT C2E}) of segmentation of the proposed methods shows a small difference between target and predicted segmentation. 


\begin{table}[h!]
    \caption{Segmentation approaches comparison on the CholecSeg8k dataset}
    \centering
    \begin{tabular}{cccc}
    \hline
    Method                                                 & IoU                               \\ \hline
    Adaptive SAM~\cite{Paranjape2024-rf}                    & 0.64                             \\ \hline
    MedT~\cite{Valanarasu2021-xq}                           & 0.57                             \\ \hline 
    TransUNet~\cite{Chen2021-hk}                            & 0.62                             \\ \hline 
    DeepLabV3+~\cite{Chen2018-at}                           & 0.56                             \\ \hline 
    UNetR~\cite{Hatamizadeh2022-rq}                         & 0.55                             \\ \hline
    DynUnet~\cite{Isensee2021-ey}                           & 0.55                            \\ \hline 
    U-Net++~\cite{Zhou2018-se}                              & 0.62                            \\ \hline 
    U-Net~\cite{Ronneberger2015-ir}                         & 0.48                            \\ \hline 
    SAMed~\cite{Zhang2023-ne}                           & 0.41                              \\ \hline
    Low Rank Adaptation of SAM~\cite{Hu2021-jj}             & 0.65                            \\ \hline 
    
    S-SAM~\cite{Paranjape2024-yj}                           & 0.71                              \\ \hline 
    SegDepth~\cite{Jamal2024-rv}                            & 0.556                             \\ \hline 
    DDA~\cite{Zhou2024-qw}                                  & 0.59                             \\ \hline 
    CLU~\cite{Sheng2024-ce}                                 & 0.463                             \\ \hline
    \textbf{C2E (ours)}                                          & \textbf{0.74}                     \\ \hline
    \end{tabular}
    
    \label{tb: sota segmentation} 
\end{table}

\subsection{Intraoperative polyp diagnosis} \label{section: results, polyps}
Detection and diagnosis of microtumor and polyps instruct surgeons and physicians to have early detection of cancer, reduce colorectal cancer mortality, reduce the cost of treatment with the costly treatment required for advanced cancer, and improve patient outcomes. Specifically, detection and diagnosis in colonoscopy can help to take the right decision on if they need to dissect the polys for further pathology tests in tumors and if they need to resect and dissect polyps attached to the colon. We adoped the PolypDiag data set~\cite{tian2022contrastive} for evaluation. For the fine-tuning settings, we used linear layer as decoder. The results have shown improvement in this task compared to several related works. 
\begin{table}[ht]
    \centering
    \caption{Approach comparison on Polyp diagnosis}
    \begin{tabular}{cccc}
    \hline
    Method                              &  Accuracy[\%]      \\ \hline 
    ProViCo[26]\cite{park2022probabilistic} & 86.9±0.5          \\\hline
    VCL~\cite{qian2022static}           & 87.6±0.6              \\\hline
    ST-Adapter~\cite{pan2022parameter}  & 84.8±0.7              \\\hline
    EndoSSL ~\cite{Hirsch2023-vl}       & 91.3±0.6              \\\hline
    EndoFM~\cite{Wang2023-nc}           & 90.7±0.4             \\ \hline
    \textbf{C2E (ours)}                        & \textbf{94.0±0.7}               \\ \hline 
    \end{tabular}
    \label{tb: sota polyp diagnosis} 
\end{table}

\subsection{Few-shot Analysis} \label{section: resutls, analysis, tsne, few-short}
Several questions arise when comparing our approach to other self-supervised visual foundation models. We give an example of the proposed method by comparing it to a VIT model~\cite{Dosovitskiy2020-dp}, such as the one used in \cite{batic2024endovit}, showing improved performance using the same datasets (see Tab.~\ref{tb: surgical datasets sources}) for our dataset content, significantly expanding on~\cite{batic2024endovit}). We compare the models in a variety of downstream tasks -- those of phase recognition, action prediction, and segmentation, and obtain superior results, as shown in Tab.~\ref{tb: phase, action triplet, IOU ablation model vit c2e}). 

\begin{table}[h!]
    \centering
    \caption{Ablation of model with datasets in laparoscopy cholecystectomy}
    \begin{tabular}{c|ccccc}
    \hline
    model                     & Acc(ele)           & $mAP_{ivt}$            & IOU                       &Acc                 \\ \hline
                              &  Phase             & Action triplet         & Segmentation              &Diagnosis                          \\\hline
    VIT                       & 89.7 $\pm$0.5      & 40.3 $\pm$ 2.2         & 73.5 $\pm$0.3             &92.5$\pm$2.2             \\ \hline
    \textbf{C2E(ours)}        & \textbf{92.5±6.9}  & \textbf{41.8$\pm$0.9}  & \textbf{74.2 $\pm$0.3}    &\textbf{94.0±0.7}       \\\hline 
    \end{tabular}
    
    \label{tb: phase, action triplet, IOU ablation model vit c2e} 
\end{table}

We note that the ability to facilitate few-shot learning is important, especially when we wish to extend surgical AI to novel procedures, which may not have many labeled procedures and yet possess a great potential to improve patient care and surgical results, such as~\cite{Eckhoff2023-wv}.
As an additional test for the foundation model property, we test explicitly the few-shot generalization of the proposed approach. We adapted HeiCo datasets for phase classification in a few-short setting, including proctocolectomy (10 videos), rectal resection (10 videos), and sigmoid resection (10 videos). Our approach enables better few-shot capability at the limit of few procedures, as shown in Tab.~\ref{tb: phase few short 8 video},~\ref{tb: phase few short 2 video}. 

\begin{table}[h!]
    \centering
    \caption{Comparison of few-shot phase recognition, 8 videos. }
    \begin{tabular}{ccccc}
    \hline
    Proc. & Lap.&  Proctoc-  & Rectal  &  Sigmoid  \\
          &  Chole.&  olectomy &  resect. &  resect.  \\
    \hline
    Method          & 8 vid.                        & 8 vid.                 & 8 vid.                & 8 vid.                       \\ \hline  
    VIT             & 75.84$\pm$2.76                & 58.8 $\pm$11.          & 60.2$\pm$ 0.1         & 60.2$\pm$ 0.1            \\ \hline 
    \textbf{C2E}    & \textbf{85.7$\pm$0.37}        & \textbf{84.6$\pm$2.2}  & \textbf{79.3$\pm$8.7} &\textbf{76.2$\pm$5.3}     \\ \hline
    \end{tabular}
    \label{tb: phase few short 8 video} 
\end{table}

\begin{table}[h!]
    \centering
    \caption{Comparison of few-shot phase recognition, 2 videos.}
    \begin{tabular}{ccccc}
    \hline
    Proc. & Lap.&  Proctoc- &  Rectal  &  Sigmoid  \\
     & Chole.&  olectomy &  resect. &  resect.  \\
    \hline
    Method       & 2 vid.                 & 2 vid.                  & 2 vid.                    & 2 vid.             \\ \hline  
    VIT          &67.16$\pm$3.00          &44.06$\pm$0.19           & 58.57$\pm$3.80            & 58.57$\pm$3.80      \\ \hline 
    \textbf{C2E} &\textbf{78.21$\pm$3.90} &\textbf{61.24$\pm$0.13}  &\textbf{67.34$\pm$0.01}    &\textbf{64.33$\pm$5.92}     \\ \hline
    \end{tabular}
    
    \label{tb: phase few short 2 video} 
\end{table}

\subsection{Representation Insights} \label{section: tsne}
\begin{figure}[h!]
    \centering

\begin{minipage}{0.48\linewidth}
\centering
    \includegraphics[width=\linewidth]{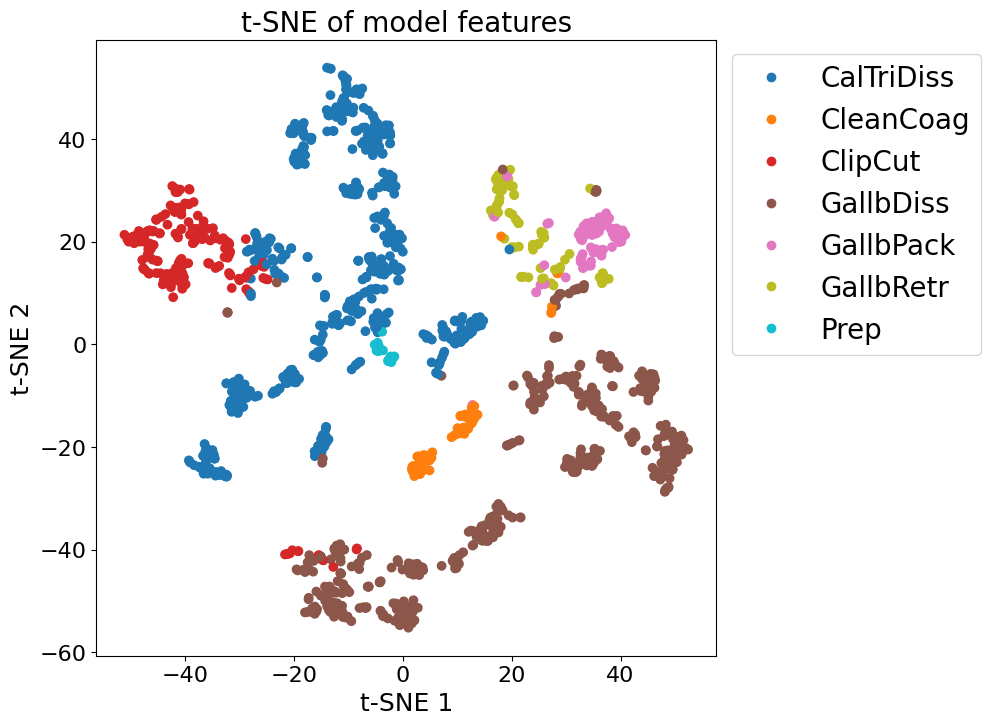}
a
\end{minipage}    
\hfill
\begin{minipage}{0.48\linewidth}
\centering
    \includegraphics[width=\linewidth]{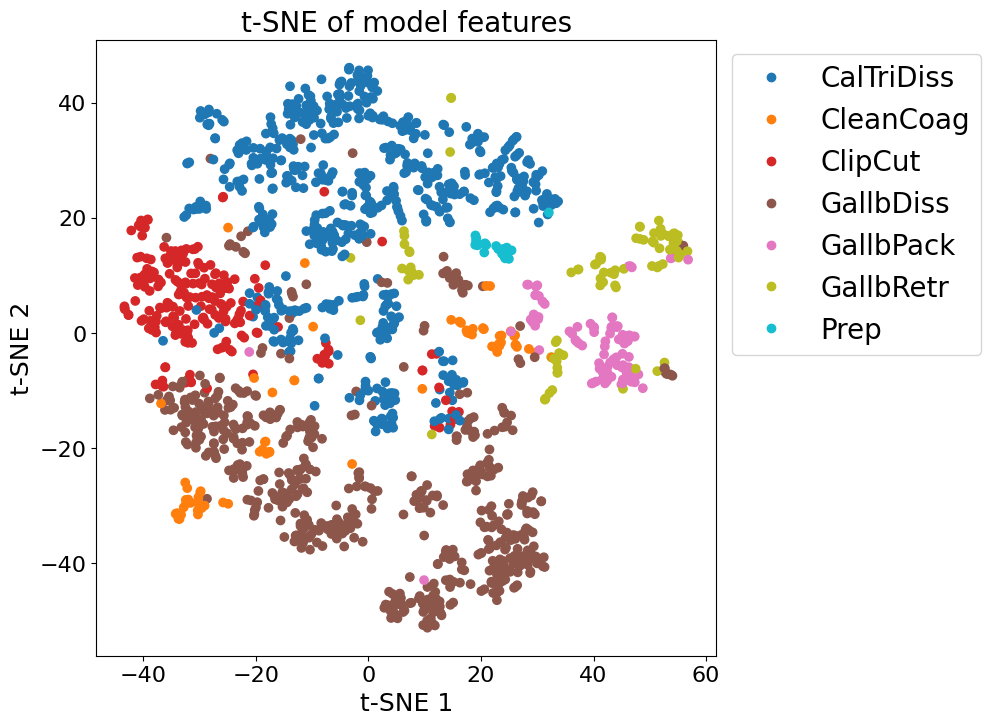}
    
    b
\end{minipage}   

    \caption{T-SNE feature visualization of our C2E approach (a) vs MAE (b), for phases from Laparoscopic Cholecystectomy procedures. Each phase is mapped to a more seperated clustered group in the embedded space compared to the entangled clusters in MAE.}
    \label{fig:phase tsne visualzation cholec80}
\end{figure}

To examine what representations are learned by the model and how well they capture and disentangle relevant surgical factors, compared to standard visual models, we visualize the two-dimensional t-SNE figure using various classes such as different types of surgery (Fig.~\ref{fig: tsne visualzation, all images}), phases of laparoscopic cholecystectomy (Fig.~\ref{fig:phase tsne visualzation cholec80}), and instruments of laparoscopic cholecystectomy (Fig.~\ref{fig:instruction, cholec80 tsne visualzation}). The features of the proposed methods are more separated compared to those obtained by probing the features obtained from \cite{batic2024endovit}. This explains why it shows promise in the few-short learning and capability of generalization.

\begin{figure}[h!]
    \centering

\begin{minipage}{0.48 \linewidth}
\centering
    \includegraphics[width=\linewidth]{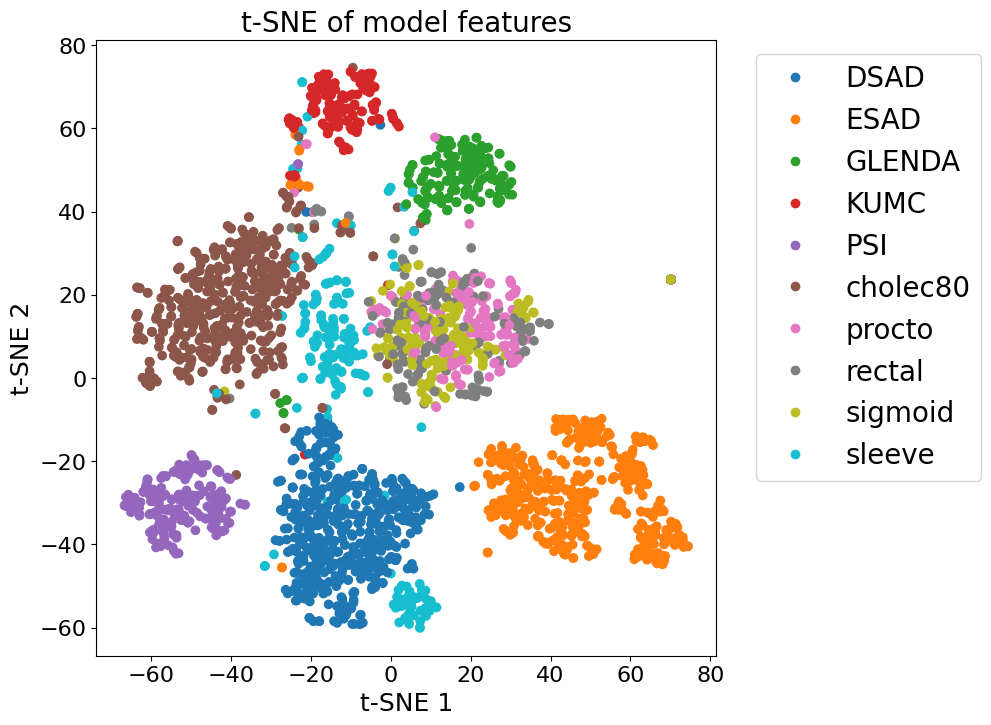}
a
\end{minipage}    
\hfill
\begin{minipage}{0.48 \linewidth}
\centering
    \includegraphics[width=\linewidth]{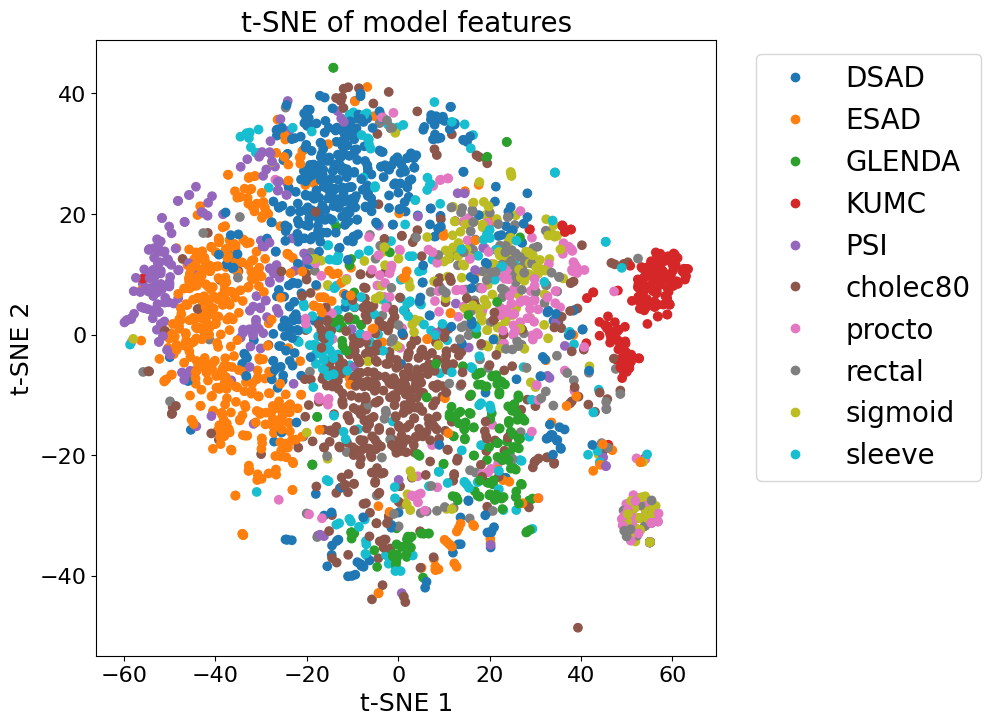}
b
\end{minipage}   
    \caption{T-SNE feature visualization of our C2E approach (a) vs MAE (b) for frames from a variety of procedures. Notice how each procedure is mapped to a different cluster in the embedded space, in a more pronounced way, compared to the entangled clusters in MAE.}
    \label{fig: tsne visualzation, all images}
\end{figure}

\begin{figure}[h!]
    \centering

\begin{minipage}{0.48\linewidth}
\centering
    \includegraphics[width=\linewidth]{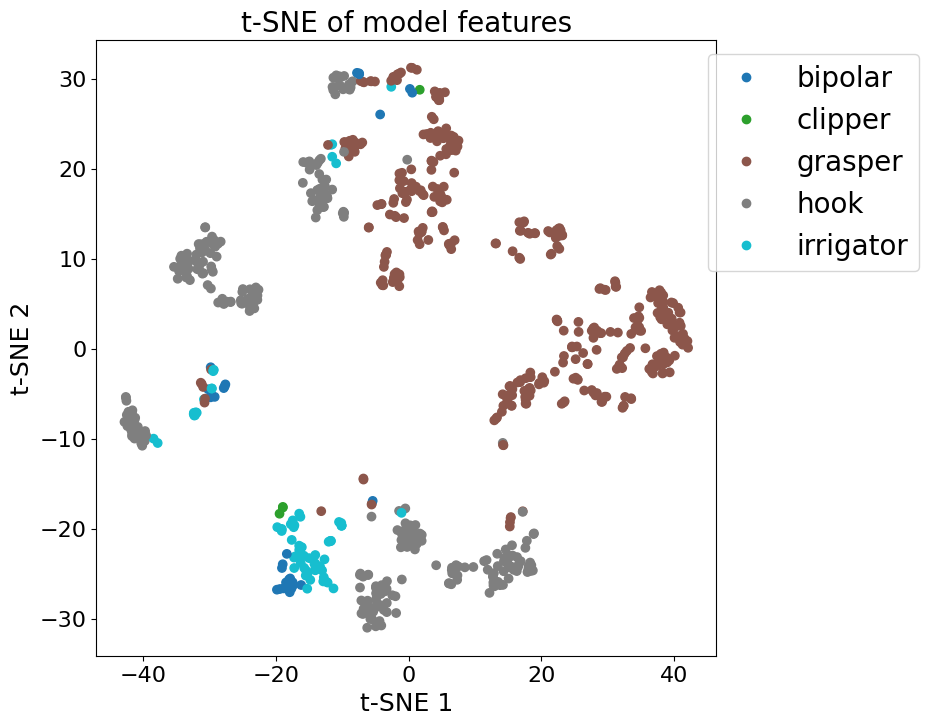}
a
\end{minipage}    
\hfill
\begin{minipage}{0.48\linewidth}
\centering
    \includegraphics[width=\linewidth]{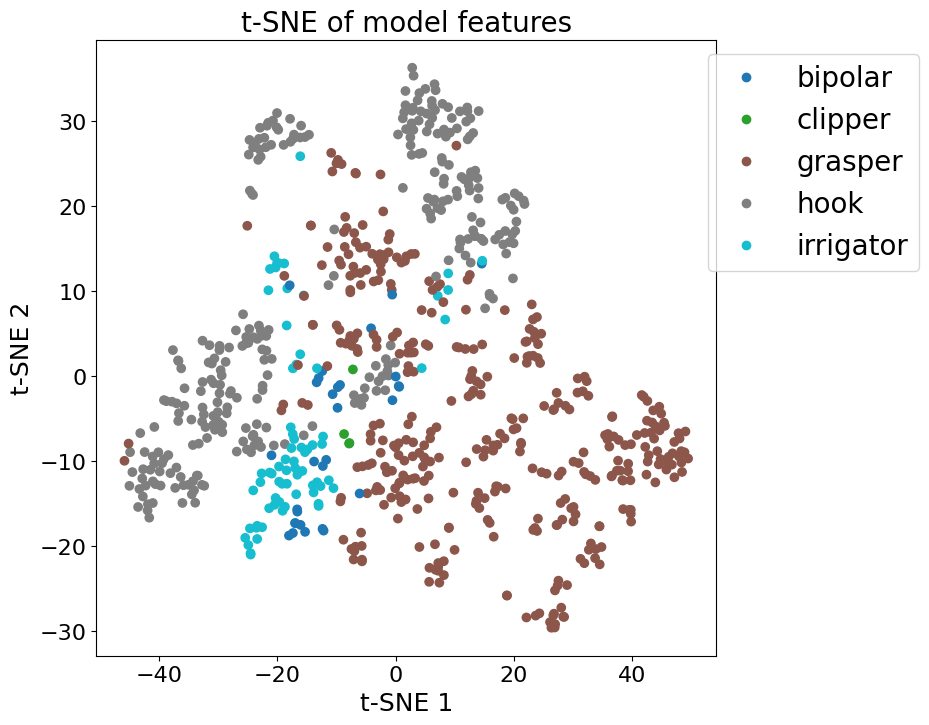}
b
\end{minipage}   
    \caption{T-SNE feature visualization of our C2E approach (a) vs MAE (b), for frames from Laparoscopic Cholecystectomy procedures. The embeddings of different instruments are disentangled in reduced dimensions.}
    \label{fig:instruction, cholec80 tsne visualzation}
\end{figure}

\begin{figure}[h!]
    \centering
\begin{minipage}{0.3\linewidth}
\centering
    \includegraphics[width=\linewidth]{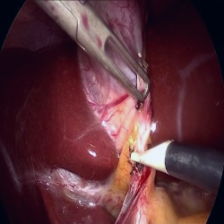}
Raw image
\end{minipage}    
\hfill
\hfill
\hfill
\hfill
\vfill
\begin{minipage}{0.3\linewidth}
\centering
\includegraphics[width=\linewidth]{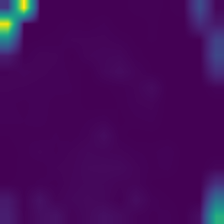}
C2E head 1
\end{minipage}   
\hfill
\begin{minipage}{0.3\linewidth}
\centering
\includegraphics[width=\linewidth]{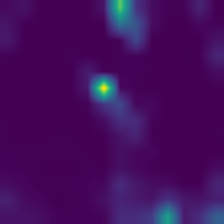}
C2E head 3
\end{minipage}   
\hfill
\begin{minipage}{0.3\linewidth}
\centering
\includegraphics[width=\linewidth]{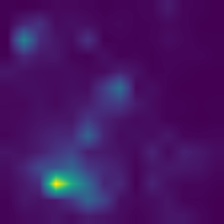}
C2E head 5
\end{minipage} 
\begin{minipage}{0.3\linewidth}
\centering
\includegraphics[width=\linewidth]{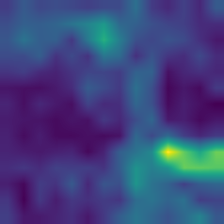}
VIT head 1
\end{minipage}   
\hfill
\begin{minipage}{0.3\linewidth}
\centering
\includegraphics[width=\linewidth]{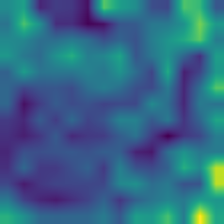}
VIT head 3
\end{minipage}   
\hfill
\begin{minipage}{0.3\linewidth}
\centering
\includegraphics[width=\linewidth]{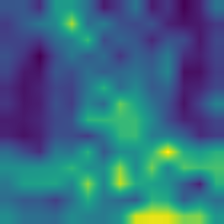}
VIT head 5
\end{minipage}   
\hfill
\begin{minipage}{0.3\linewidth}
\centering
\includegraphics[width=\linewidth]{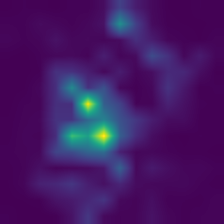}
C2E head 7
\end{minipage}   
\hfill
\begin{minipage}{0.3\linewidth}
\centering
\includegraphics[width=\linewidth]{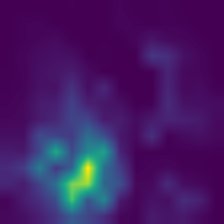}
C2E head 9
\end{minipage}   
\hfill
\begin{minipage}{0.3\linewidth}
\centering
\includegraphics[width=\linewidth]{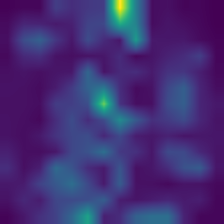}
C2E head 11
\end{minipage} 
\begin{minipage}{0.3\linewidth}
\centering
\includegraphics[width=\linewidth]{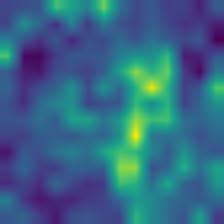}
VIT head 7
\end{minipage}   
\hfill
\begin{minipage}{0.3\linewidth}
\centering
\includegraphics[width=\linewidth]{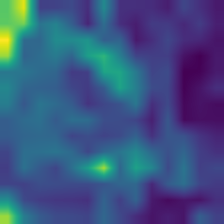}
VIT head 9
\end{minipage}   
\hfill
\begin{minipage}{0.3\linewidth}
\centering
\includegraphics[width=\linewidth]{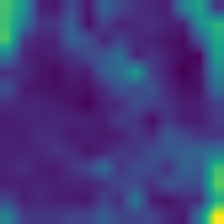}
VIT head 11
\end{minipage}   
\caption{Saliency map comparison of C2E and VIT head. The propsed method C2E has a relative sparsity of the activation maps indicating a more compressed representation. The high attention region of each heads with C2E are complementary and not overlapped indicating that the information of images are decomposed into different local information in each heads and represent global by aggration.}
\label{fig:saliency comparision  cholec 380}
\end{figure}

We visualize the saliency maps of Fig.~\ref{fig:saliency comparision  cholec 380} to provide more insights of representation of the neural networks. Compared to VIT based encoder, the saliency maps are more compact. More importantly, the peak of attention of each head differs from each other. This attention distribution demonstrated that each head represents specific local features and details information and the aggregation of multi-heads represents the global information, which creates a better balance of representation of local and global information.

\section{Conclusion}
In conclusion, we leverage the compression method and entropy maximization to encourage the compression of images and improve the representation of neural network in pertaining. The pre-trained backbone has shown promising performances in various downstream tasks in the surgical artificial intelligence domain, such as workflow prediction, robotic action, segmentation, and diagnosis. We envision that the proposed self-supervised learned foundation model can reduce the labeling quantities required for various downstream tasks since high-quality annotation in surgical videos is limited due to privacy and high working load. It will enables an improved quality of various of monitoring tasks and reduce the complication of surgery, and improve quality of treatment.  


\section{Main theoretical contribution}

\begin{theorem} \label{theorem: min H min Kolmogorov complexity}
The entropy of the image embeddings $H(X)$ is the Kolmogorov complexity $K(X)$ of image embeddings in compression.
\end{theorem} 
\begin{proof}
Based on Theorem 14.3.1 \cite{Cover2006-se}

\begin{equation}
\begin{aligned}
H(x) \leq \mathbb{E}\frac{1}{n}K(X) \leq H(X) + \frac{(|\chi|-1 )\log{n} + c }{n} \\
\end{aligned}
\end{equation}
\begin{equation}
\mathbb{E} \frac{1}{n} K(X) \rightarrow  H (X)
\end{equation}

This equation holds on condition of $x \in \chi$ is finite. The domain of input information of the images is finite. The transformation of input to the hidden state is also finite. 

Therefore, the Kolmogorov complexity $K(X)$ approaches the entropy $H(x)$ for large $n$. 

\end{proof}

\begin{theorem} \label{theorem: convex of maximize compression}
     The compression function Eq.~\ref{eq: maximize compression = maximize entropy difference} is convex
\end{theorem}
\begin{proof}
Combine Eq.~\ref{eq: Kolmogorov complexity} and Eq.~\ref{eq: maximize compression = maximize entropy difference}
\begin{equation}
    \begin{aligned}
         \quad & - H(Z_{i+1}) + H(I_{i+1}) \\
        =&  -\frac{1}{2} \ln{|\Sigma|} - \frac{D}{2} (1 + \ln{2\pi}) + H(I_{i+1})
    \end{aligned}
\end{equation}
Based on Theorem 7.6.7 in \cite{johnson1985matrix} the function $\ln{|A|}$ is a strictly concave function on the convex set of positive definite Hermitian matrices.

While $\Sigma$ is positive definite Hermitian, therefore, $\ln{|\Sigma|}$ is concave.

The $-\frac{1}{2} \ln{|\Sigma|}$ is convex. The solution of maximization is located on the edges of the scope of hidden space. 
\end{proof}

\begin{theorem} \label{theorem:  z = z_k -(-H + H) minimize Kolmogorov complexity}
The gradient step Eq.~\ref{eq: z = z_k -(-H + H)} minimizes Kolmogorov complexity. 
\end{theorem}
\begin{proof}
Consider constant $n$ during training.
Equation ~\ref{eq: z = z_k -(-H + H)} minimizes $H(Z)$ which is a higher bound of $K(X)$. Therefore, optimization minimizes Kolmogorov complexity.  
\end{proof}

\begin{theorem}
\label{theorem: dimension reduction process}
The stepwise dimension reduction process in Eq.~\ref{eq: entropy of multivariable Gaussion} is equivalent to maximizing the compression step in Eq.~\ref{eq: maximize compression = maximize entropy difference}.

\end{theorem}
\begin{proof}
\begin{equation} \label{eq: dim expansion, maximization} 
    \begin{aligned}
        &(H(Z_{i}) - H(Z_{i+\frac{1}{2}})) \\
        = & \frac{1}{2}(\ln{|Z_{i}|} - \ln{|Z_{i+1}|} - |z_{i}|\nabla_{Z_{i+1}}(\ln{|Z_{i+1}|}) ) \\
        = & \frac{1}{2}(\ln{|Z_{i}|} - \ln{|Z_{i+1}|} - |z_{i}| |Z_{i + 1}|^{-1} )
    \end{aligned}
\end{equation}

where, $Z_{i+\frac{1}{2}}$ is the hidden state without dimension reduction. From first to second equation use Taylor expansion and the $|Z_{i}|>> |z_{i+1}|$ and the dimensionality of $dim(z_{i}) << dim(Z_{i + 1})$, therefore the following equation holds
\begin{equation}
    \begin{aligned}
        \arg\max_{Z_{i+\frac{1}{2}}} & H(Z_{i}) - H(Z_{i+\frac{1}{2}}) \\
        = & \arg\max_{Z_{i+1}}\frac{1}{2}(\ln{|Z_{i}|} - \ln{|Z_{i+1}|})  \\
    \end{aligned}
\end{equation}
\end{proof}

\begin{theorem} \label{theorem: maping Z to subpsace maximize the compression}
The mapping of embedding $Z$ to an orthogonal subspace maximizes the Kolmogorov complexity difference defined as $- H(Z) + H(I)$
\end{theorem}

\begin{proof}
Since $\ln{|Z|}$ is concave, $Z$ is a positive definite Hermitian matrix. 

\begin{equation}
     \begin{aligned}
        & \max_{Z}\frac{1}{2}( -\ln{|Z|}) =   & \min_{Z}\ln(|Z|) \\
    \end{aligned}   
\end{equation}
since $\ln(Z^TZ)$ has a lower bound of $\sum_i^{k}\ln({Z^T\Pi_iZ})$ where, $\Pi \in R^{C,C}$ is a diagonal matrix with $\pi_i \in [0,1]$ and $ \sum_i^k \Pi_i$ as identity matrix. 
\begin{equation} \label{eq: minimization of Z}
    \begin{aligned}
         & \max_{Z}\frac{1}{2}( -\ln{|Z|}) =   & \min_{Z}\ln(|Z|) \\
    =   & \min_{Z,\Pi} \sum_i^{k}\ln({Z^T\Pi_iZ}) \\
    \end{aligned} 
\end{equation}
As $\ln(Z^T\Pi_i Z)$ is a concave function (Theorem 7.6.7 in \cite{johnson1985matrix}) for $Z^T \Pi_i Z$ to be positive semi-defined. The global minimization is at the boundary of the convex domain, where $\pi_i \in \{0,1\}$. This shows that $\Pi_i$ maps $Z$ to the $k$ subspace to maximize the compression. 

\end{proof}

\begin{theorem} \label{theorem: temperature sub}
The temperature $T$ defined in expectation of energy $\mathbb{E}[e(z)] = kT$ in Eq.~\ref{eq: Z_k-1/2 = z_k-1} is conditioned by partial elements of the hidden state $z$
\end{theorem}
\begin{assumption} \label{assumption: particle z in space is homogeneous}
The distribution of particles $z$ in all its subspaces is homogeneous. 
\end{assumption}
\begin{proof}
The $z$ can be split into two groups $z = [z_{sub_i},z_{sub_j}]$. Per Assumption~\ref{assumption: particle z in space is homogeneous},
\begin{equation} \label{eq: E [e(z)] = E[e(z_{sub_i})] }
\mathbb{E}[e(z)] = \mathbb{E}[e(z_{sub_i})]
\end{equation}
\end{proof}

\bibliographystyle{ieeetr}
\bibliography{references}

\end{document}